\documentclass[draft]{agujournal2019}
\usepackage{url} 
\usepackage{lineno}
\usepackage[utf8]{inputenc}
\usepackage{gensymb}
\usepackage{amsmath}
\usepackage{amsfonts}
\usepackage{amssymb}
\usepackage[T1]{fontenc}
\usepackage{graphicx}
\usepackage{subcaption}
\usepackage{physics}
\usepackage{epigraph}
\usepackage{siunitx}
\usepackage[sectionbib]{natbib}
\usepackage{chapterbib}

\draftfalse

\journalname{Geophysical Research Letters}

\begin{document}
\title{Quantifying intra-regime weather variability for energy applications}

\authors{Judith Gerighausen\affil{1}, Joshua Dorrington \affil{1,2}, Marisol Osman \affil{1,3}, Christian M. Grams \affil{1,4}}

\affiliation[1]{Institute of Meteorology and Climate Research (IMKTRO), Department Troposphere Research, Karlsruhe Institute of Technology (KIT), Karlsruhe, Germany}
\affiliation[2]{now at: Geophysical Institute (GFI), University of Bergen (UiB), Norway}
\affiliation[3]{Universidad de Buenos Aires, Facultad de Ciencias Exactas y Naturales,
Departamento de Ciencias de la Atmósfera y los Océanos. CONICET – Universidad de Buenos Aires. Centro de Investigaciones del Mar y la Atmósfera (CIMA). CNRS – IRD – CONICET – UBA. Instituto Franco-Argentino para el Estudio del Clima y sus Impactos (IRL 3351 IFAECI).}
\affiliation[4]{Federal Office of Meteorology and Climatology, MeteoSwiss, Zürich-Flughafen, Switzerland}

\correspondingauthor{Joshua Dorrington}{joshua.dorrington@uib.no}




\begin{keypoints}
\item Variability in surface weather within a particular weather regime can be large. 
\item Studying the evolution of scalar weather regime indices can provide valuable insights into intra-regime variability 
\item A refined application of weather regimes is provided with applications for stakeholders in the energy sector. 
\end{keypoints}

%
%

%
%


\begin{abstract}
Weather regimes describe the large-scale atmospheric circulation in the mid-latitudes in terms of a few circulation states that modulate regional surface weather. Subseasonal forecasts of prevailing weather regimes have proven skillful and valuable to energy applications. Previous studies have mainly focused on the mean surface weather associated with a regime. However, we show in this paper that variability of surface weather within a regime cannot be ignored. These intra-regime variations, caused by different `subflavors' of the same regime, can be captured by continuous regime indices and allow a refined application of weather regimes.
Here we discuss wintertime temperature and wind speed regime anomalies for four selected countries, and provide guidance on the operational use and interpretation of regime forecasts. In an accompanying supplementary dataset we provide similar analysis for all European countries, seasons and key energy variables, useful as an applied reference.
\end{abstract}

\section*{Plain language summary}
A large fraction of European surface weather can be explained by variation between a small handful of `weather regimes': continental-scale circulation patterns which change the likelihood of warm and cold / windy and calm conditions. Because these large-scale regimes are predictable on longer time scales they have proven useful for energy applications. But previous work has only focused on the `typical' weather associated with a regime, ignoring the prospect of e.g. warm conditions in a normally cold regime. We show that this can be misleading for applications, and provide advice on making better use of regime forecast information.
%
%


\section{Introduction}
The direct prediction of surface weather in Europe with any deterministic accuracy has long been limited to between 1 and 2 weeks; a result of the chaotic nature of the atmosphere. However, an ever-escalating demand from stakeholders has emerged for guidance on the subseasonal timescale, spanning  from 2 to 6 weeks ahead. The energy sector is one of the largest of these stakeholders, as an accurate understanding of power demand (driven predominately by temperature) and renewable power output (impacted by near-surface wind and solar radiation) are vital for, e.g. forward planning of grid and power plant maintenance and hedging energy futures to maintain price stability.
Fortunately, the Euro-Atlantic atmosphere features considerable low-frequency variability in the form of weather regimes: large-scale flow configurations that capture such structures as blocking anticyclones and jet stream deviations. Forecasts of a prevailing weather regime are inherently probabilistic, exclude outcomes that are virtually impossible in certain flow configurations and emphasize highly favored outcomes. Forecasts of the prevailing regime in extended-to-subseasonal timescales have proven skillful (\cite{Ferranti_2015, matsueda_estimates_2018, bueler_yearround_2021, osman_multimodel_2023}) and useful for the energy sector (\cite{van_der_wiel_influence_2019, Bloomfield_WR_TCT_2020}) and the impact of regimes on surface weather and renewable energy has been well explored (e.g. \cite{grams_balancing_2017, liu_how_2023, Mockert_2023}).
However, except for \citet[the supplement of][]{grams_balancing_2017} and the discussion of flow-dependent surface weather predictability in \citet{Spaeth_2024} little consideration has so far been given to how surface weather varies \emph{within} a given regime. While this may seem a minor oversight at first glance (after all, is the purpose of weather regimes not to simplify the continuous variation of weather into a manageable number of discrete mean-states?) we will show in this paper that, the variance of surface weather within a regime should also be considered. We also demonstrate how distinct 'flavours' of a regime -- with radically different surface impacts -- can be simply identified using continuous regime indices.

Specifically, we adopt the seven regime framework of \cite{grams_balancing_2017}, and present a detailed analysis of intra-regime variability in \SI{2}{m} temperature (T2m) and \SI{100}{m} wind speed anomalies (W100m) during the winter (DJF) season, when European energy demand is highest. Walking through a small number of demonstrative examples, we highlight the importance of intra-regime variability and how, in many cases, it can be accounted for to distinguish between cold/calm and warm/windy anomalies in a given regime. An extensive Supplementary Dataset \citep{Zenodo_2024} provides a more comprehensive analysis of regime anomalies in all seasons for T2m, wind speed, total precipitation and insolation anomalies for all countries in the European Economic Area (as of 2024), Switzerland, and the United Kingdom (UK). 

We describe our data and methods in Section 2, analyse our results in Section 3, and present our conclusions in Section 4.

\section{Data and methods}
We use the ERA5 reanalysis from the European Centre For Medium-Range Weather Forecasts \cite{hersbach_era5_2020} over the 43-year period from 01/01/1979--31/12/2021. 
\subsection{Weather regime definition}
Seven year-round weather regimes are defined following \cite{grams_balancing_2017}, based on EOF-analysis and k-means clustering of 10-day low-pass filtered 500 \si{\hecto\pascal} geopotential height anomalies (Z500') in the North Euroatlantic sector (80\degree W-40\degree E , 30\degree N-90\degree N), adapted to ERA5 as explained in \cite{Hauser:2023}. The weather regime index, ($I_{wr}$), serves as a continuous measure of weather regime activity and is defined as the normalized projection of 3-hourly Z500' onto each of the seven cluster means. As well as being useful for characterising North-Euroatlantic flow variability, the $I_{wr}$ also permits the definition of discrete active life cycles for each regime. Concretely, a regime, $R$, is considered active if $I_{wr=R}>1.0$ for at least five consecutive days. If no regime meets this criterion for a given date, that date is categorized as "no regime". 

\subsection{Surface weather anomalies}
Daily mean anomalies (computed from 6-hourly ERA5 data) of T2m and W100m are computed with respect to the 31-day running mean climatology at a given calendar day (1979-2021) using data on a 1\degree grid. Then, a linear trend is removed to account for climate change and a five-day rolling mean is applied to remove sub-synoptic variability.\\

The surface weather anomalies are spatially-averaged over the national boundaries of European countries (masked with Python-package "regionmask"), using cosine-latitude weighting, to yield scalar time-series of country-aggregated weather. 
These time series are stratified into terciles for a given regime and season to investigate differences between cold/calm and warm/windy situations. We note that stratifying by surface extremes (i.e. the 5th and 95th percentiles) provides qualitatively similar results, albeit with reduced statistical robustness \citep[][retrieved at 7 July 2024]{Gerighausen:2024}. After stratification into terciles, we consider the averaged temporal evolution of all seven $I_{wr}$ indices in a 20-day window centred on each date in the respective tercile. The resulting lagged $I_{wr}$ composites allow us to see if an upper or lower tercile event systematically occurs in a late or early stage of a regime's life cycle, or if other regimes are simultaneously active. Significant differences in intra-regime flow evolution are quantified with a Kolmogorov-Smirnov test at the 1\% level between $I_{wr}$ indices in the tercile of interest and those in the other two terciles at each time step.

\subsection{Metrics}
To analyze how surface weather anomalies vary within days belonging to a given regime, we define two gridpoint metrics: the fractional standard deviation (FSTD), $\sigma_{frac, season}$:
\begin{align}
    \sigma_{frac, season}=\frac{\sigma_{regime, season}}{\sigma_{season}}.
\end{align}
and the signal-to-noise ratio (S2N):
\begin{align}
    S2N=|\frac{X_{regime, season}}{\sigma_{season}}|,
\end{align}
where $X$ stands for the T2m or W100m anomaly. 

These metrics simply correspond to the standard deviation and the (sign-invariant) mean of surface weather anomalies in a regime respectively, scaled by the seasonal standard deviation of surface weather. Values of $FSTD < 1$ indicate a reduced range of surface-weather variability within a regime than in the season as a whole, with $FSTD = 0$ corresponding to the naive perspective that \emph{all} surface weather is accounted for by regime activity. Conversely, $FSTD>1$ indicates surface weather is particularly uncertain within a given regime, due to sensitivity to small-changes in the large-scale flow.

High S2N, meanwhile, indicates that surface weather anomalies are meaningfully strong compared to climatic variability. If S2N is low, then the net shift in surface weather is negligible in comparison to variability. Importantly however, low S2N does not necessarily imply no impact of the given regime on surface weather. This is only the case if FSTD is close to 1.

\section{Results}
\subsection{Modulation of surface weather variability by weather regimes in winter}

\begin{figure}[h]
    \centering

    \includegraphics[width=\linewidth]{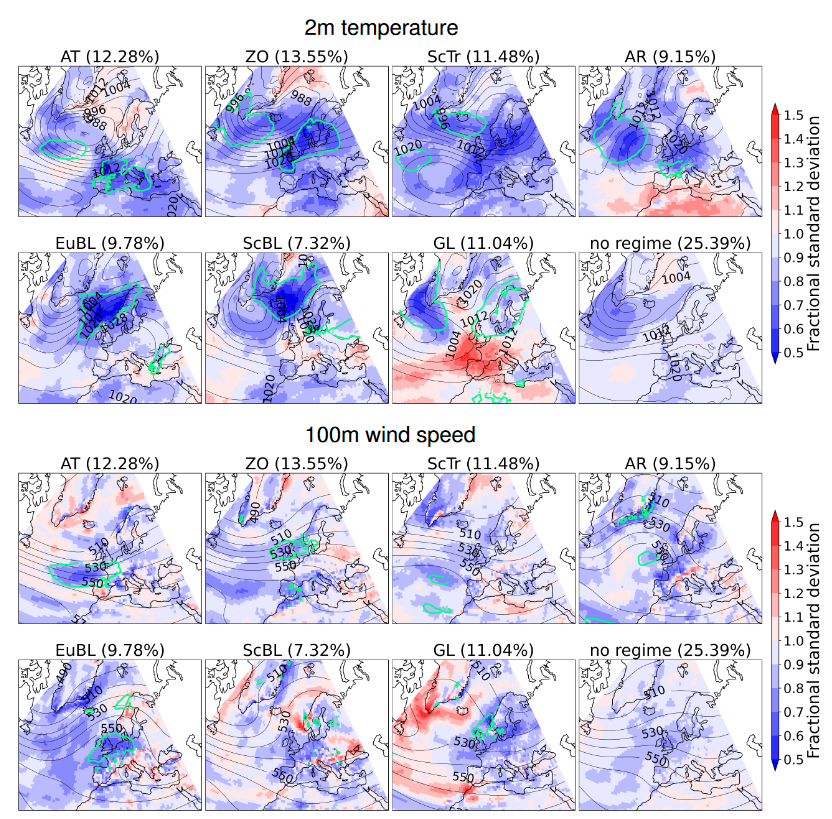}\\
    \caption{Upper Rows: Fractional standard deviation (FSTD) of \SI{2}{m} wintertime temperature regime anomalies with mean sea level pressure overlaid (contours every 4\,hPa). Lower Rows: FSTD of \SI{100}{m} wind speed  anomalies with 500\,hPa geopotential height overlaid (contours every 10 gpdm). Cyan contours indicate areas with signal to noise ratio >1 (cf. Supplemental Figures S2 and S4). Percentages above each panel indicate wintertime regime occurrence frequency.}
    \label{FSTD}
\end{figure}

Broadly, wintertime regimes can be split into two categories: the cyclonic regimes, Atlantic Trough (AT), Zonal (ZO), and Scandinavian Trough (ScTr) regimes which bring mild and windy conditions, and the anticyclonic regimes, Atlantic Ridge (AR), European Blocking (EuBL), Scandinavian Blocking (ScBL), and Greenland Blocking (GL) bringing calm but colder weather to Central Europe especially. Within this broad categorisation there is a large variety of regional variation. These mean impacts have been discussed in detail in \citep{grams_balancing_2017} and are not discussed further here, although temperature and wind anomalies are shown in Supplemental Figures S1 and S3. Instead we focus on variability of surface weather within each regime.\\
In general, DJF T2m and wind variability is reduced within a given regime compared to climatology (i.e. FSTD<1), as we would hope, indicating the practical utility of regimes for modulating surface weather (Figure \ref{FSTD}). For T2m the notable exceptions are Mediterranean temperatures during AR, and -- particularly clearly -- south-west European temperatures during GL, which both show greater than climatological variability. For GL, this region of uncertain temperatures is co-located with a southerly upper level jet and the eastern edge of a surface cyclone in the composite mean (c.f. Z500 contours and MSLP contours in \ref{FSTD}, 4th and 2nd row, respectively), reflecting the southward shift of the storm track during GL \citep[cf.][]{Hauser:2023a}. In this region, mean temperature anomalies are weak (Figure S1) and S2N is low (cyan contour in Figure \ref{FSTD}, and Figure S2): there is little mean effect of GL on southwest European temperature. It is only over Northern Europe where FSTD<1 and S2N>1, indicating reliably and robustly colder weather than usual. Elsewhere the mean impact of GL is uncertain.

Intra-regime W100m variability is comparatively more heterogeneous (Figure \ref{FSTD}, 3rd and 4th row). Areas of high S2N are rare and spatially localised. Importantly, during AT and ZO, high S2N together with low FSTD occur over the western coasts of Europe and the North Sea, respectively, colocated with robust positive W100m anomaly during AT and ZO (cf. Figure S3), which has relevance for regional wind power production \citep[cf.][]{grams_balancing_2017}. These areas with low FSTD and high S2N indicate regime configurations that provide windows of predictive opportunity, where surface dynamics are well-controlled by the large-scale regime. Overall during all zonal regimes (AT, ZO, ScTr) intra-regime variability along western and northern European Coasts is reduced. However, further inland variability is also enhanced while S2N is generally low.\\
During blocked regimes (AR, EuBL, ScBL, GL) wind speed varies less in the core (AR, EuBL, ScBL) or in the case of GL at the eastern flank of the accompanying upper-level ridge and surface high (contours in Figure \ref{FSTD}). As for the zonal regimes, local areas can see enhanced intra-regime variability, such as the Alps and parts of Scandinavia and Eastern Europe for AR, EuBL, and ScBL. GL sees enhanced wind speed variability over the North Atlantic and southern Spain. Notably, a high S2N over the North Sea region in EuBL and GL indicates robust calm conditions (cf. Figure S4) in alignment with \citet[cf.][]{grams_balancing_2017, Mockert_2023}. 

\newpage
\subsection{Details on intra-regime temperature variability}
\begin{figure}[h]
\centering
    \includegraphics[width=\linewidth]{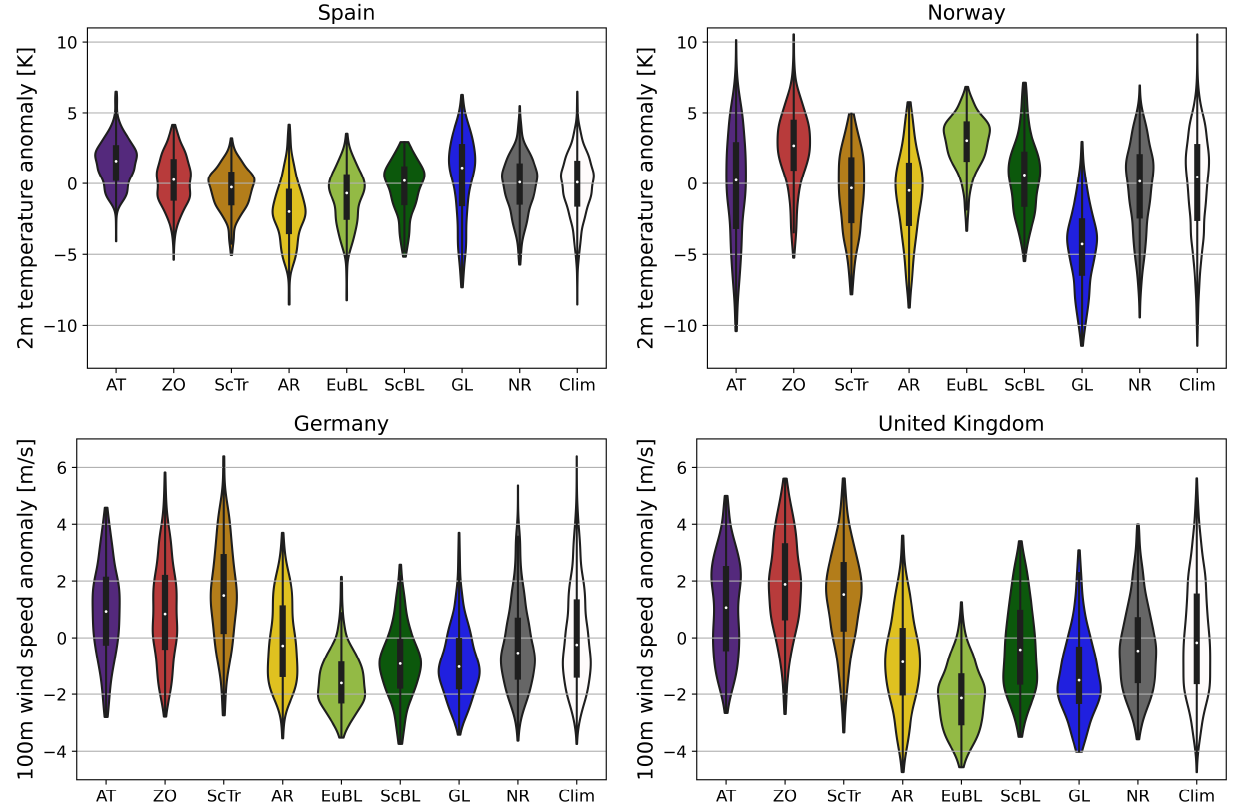}\\
\caption{Probability distributions of \SI{2}{m} temperature anomalies (T2m) in Spain and Norway (upper row) and \SI{100}{m} wind speed anomalies (W100m) in Germany and the United Kingdom (lower row) for each regime, no regime and the wintertime climatology.}
\label{fig:violins}
\end{figure}

More insights into intra-regime variability in specific regions can be gained by using the distributions of country-aggregated surface weather anomalies (Figure \ref{fig:violins}). As examples, we discuss intra-regime T2m variability for Spain and Norway to shed light on the marked FSTD pattern for GL discussed in the previous section. In Section 3.3 we then discuss intra-regime W100m variability for the UK and Germany, motivated by the high share of wind power in their electricity systems. Similar analysis and plots of intra-regime variability  for all countries, variables (T2m, W100m, precipitation, insolation), and seasons are provided as a supplementary dataset. 
Wintertime temperature variability in Spain is largest during GL spanning a range of approximately \SI{14}{K} (c.f. Figure \ref{fig:violins}, Spain); comparable to the total climatology. With a median of approximately \SI{1}{K}, GL is the second warmest regime in Spain on average, but features a long cold tail. As such, realisable cold extremes during GL are among the three coldest regimes after AR and EuBL. 
This non-negligible probability of cold anomalies is remarkable and totally missing from the mean picture, which views GL as mildl in Iberia. This could easily lead to false forecasting assumptions.\\

In contrast, GL brings reliably cold weather to Norway in winter, with a median temperature of about \SI{-4}{K} and only a small fraction of days above \SI{0}{K}. Thus, the variability of GL in this region defines the degree of negative anomalies; whether it will be cold or extremely cold.\\

\begin{figure}
    \centering
    \includegraphics[width=\linewidth]{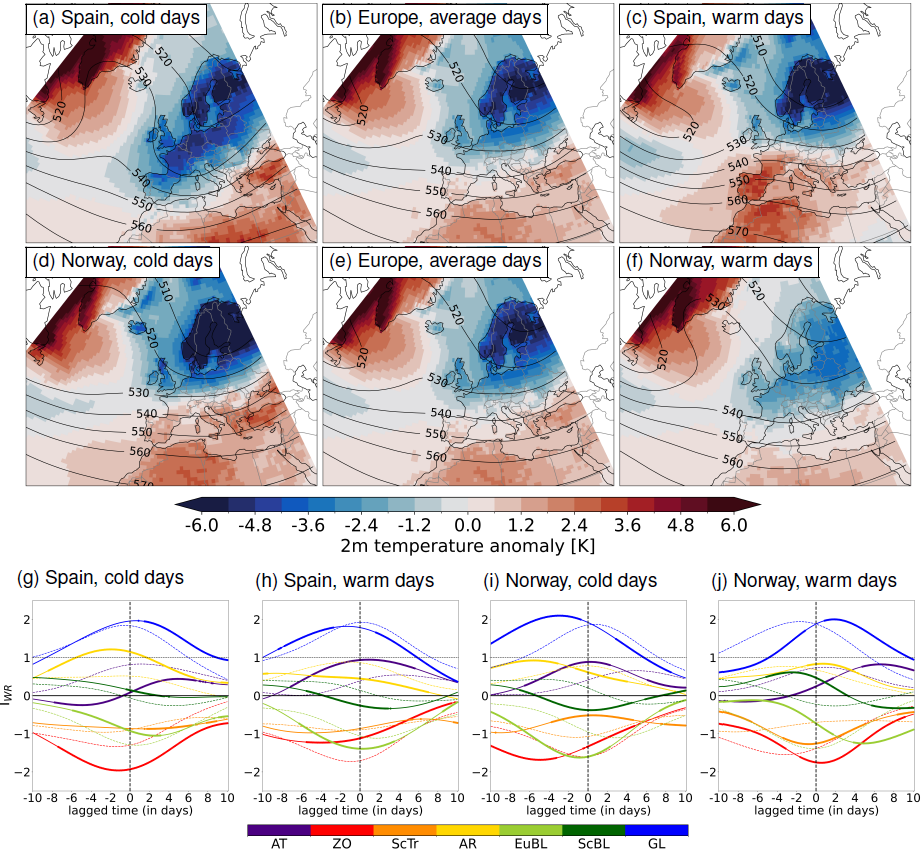}
    \caption{Composites of wintertime \SI{2}{m} temperature (T2m) anomalies with absolute geopotential height composites overlaid for Greenland Blocking (GL) days exhibiting lower (a), d)) and upper (c),f)) tercile T2m anomalies in Spain (a), c)) and in Norway (d), f)). Composites for all wintertime GL days are displayed in b) and e). Panels g)-j): Lagged composites of the $I_{wr}$ index in a 20-day window centered on the date of lower/upper tercile events in both regions. Significant differences (Kolmogorov-Smirnov test of 1\%) between the tercile of interest (solid line) and the other terciles (dashed lines) are marked in bold.}
    \label{fig:temp_example}
\end{figure}

In order to give guidance in these two contrasting cases of Spain and Norway, we explore differences in the atmospheric flow pattern for GL days in the upper and lower tercile of T2m anomalies (Figure \ref{fig:temp_example} a-f). 
When GL is cold over the Iberian Peninsula, the ridge is shifted east over the Atlantic Ocean compared to average GL days (Figure \ref{fig:temp_example} b), and the trough reaches from Scandinavia deep into Central Europe (Figure \ref{fig:temp_example} a). The jet stream shifts south over northern Africa and negative T2m anomalies extend over almost all of Europe. The $I_{wr}$ composites (Figure \ref{fig:temp_example} g) reveals that the cold GL days in Spain occur around the peak of GL life cycles and tend to be longer but not necessarily stronger events. More relevant in explaining contrasting T2m anomalies during GL in Spain is the co-projection into a secondary regime, namely AR, about three days prior to the coldest GL days in Spain. AR $I_{wr}$ first exceeds 1 at day -6, shortly after the onset of GL and decays around the peak of GL at day +2. At the same time there is a strong anti-projection into ZO. The co-occurring AR and strongly suppressed ZO reflects exactly the eastward shifted and intensified ridge-trough setup, supporting cold air advection from the North and Northeast into the continent. 
Warm GL days in Spain occur when the ridge over the Atlantic has a northwestern-southeastern tilt, and the jet is wavier. The flow pattern over the western Mediterranean is more south-westerly, leading to warm air advection, while the cold air remains confined over Scandinavia and eastern Europe (Figure \ref{fig:temp_example} c). This occurs shortly after the peak of GL together with a significantly stronger co-projection in AT, while the signature of the other anticyclonic regimes is weaker (Figure \ref{fig:temp_example} h).\\
Thus for GL days in Spain a co-projection in AR results in cold anomalies, whereas AT co-projection results in warm anomalies. A similar behaviour is observed for Western and Central European countries such as Germany (see Supplementary dataset). In particular for Germany, a co-projection in AR or AT  modulates the potential for energy-critical cold and calm conditions during GL \citep{Mockert_2023}. 

Now considering Norway, extreme cold during GL shows a ridge shifted westward (c.f. Figure \ref{fig:temp_example} panels d and e),  over the North Atlantic and Greenland. The trough over Scandinavia is more extensive than average bringing extremely cold anomalies to the whole region. Adjacent regions, such as the British Isles or Poland, also experience colder anomalies from the stronger intrusion of polar air. Southern Europe experiences modestly strengthened warm anomalies, and so temperature gradients over central Europe are intensified. Relatedly, the jet stream is shifted south and intensified over the Mediterranean (contours of 500\,hPa geopotential height in Figure \ref{fig:temp_example} d).\\
These flow changes again project well onto the $I_{wr}$ (Figure\ref{fig:temp_example} i). From a secondary regime perspective there is a relatively strong co-projection into AR prior to extreme cold days, and into AT during "warmer" days, while other regimes are suppressed, but here the main GL index is also quite different. Very cold GL events are on average two days longer than warmer GL events (in terms of period with mean $I_{wr}>1.0$) and the cold days (lag 0) generally occur in the decaying stages of the regime, with peak GL $I_{wr}$ 3-4 days prior and has higher amplitude than in the other terciles. This suggests a clear interpretation where the deep trough over Scandinavia, together with the ridge over Greenland leads to stronger northerly advection into Scandinavia, which builds to a peak during the final days of a long-lived GL event \citep{Bieli_2015}.\\
During upper-tercile T2m GL days over Norway, the anomalies are still cold, but mildly so, and spread over large parts of north and central Europe (Figure \ref{fig:temp_example} f). The ridge over the Atlantic is more pronounced and much more west-east tilted than on average, while the trough over Scandinavia is weaker. The "warm" Norwegian GL days occur in the early stages of GL life cycles, about two days prior to the maximum mean $I_{wr}$, indicating that the GL life cycle is still in the developing phase (Figure \ref{fig:temp_example} j). Interestingly, a co-projection into ScBL and AR occurs before and around the "warm" GL days in Norway, and only later a co-projection in AT evolves. Thus monitoring the projection in AR, ScBL, and AT might help to assess the degree of cold response to GL in Norway. 


In summary, variations in the atmospheric flow pattern, that nevertheless project into the GL regime, are able to explain the variability of temperature anomalies over different countries. These changes project differently into the $I_{wr}$ of the active, secondary, and suppressed regimes and reveal that in some cases, e.g. Norway, the anomalies depend strongly on the stage of the regime (early or late), while, for example, in Spain, co-occurring secondary regimes are of more importance. This knowledge can be used in weather forecasting by not only monitoring the potential most likely regime, but also the "flavour" of the current regime as captured by $I_{wr}$ co-projections.

\subsection{Wind speed variability}
The distributions of W100m anomalies show that ScTr and ZO are the windiest regimes for Germany and the UK, respectively. Nevertheless, cyclonic intra-regime variability is clearly large, and even these windiest regimes can produce wind anomalies below \SI{-2}{m/s}. The large range of variability with partially 25\% of days below average during cyclonic regimes can lead to unexpected below average wind power output.

The anticyclonic regimes are much less variable, especially EuBL, which brings calm conditions to both countries in almost all cases. AR in Germany and ScBL in the UK are calm on average as well but can bring windy conditions almost half the time. As the intra-regime variability of W100m is rather similar for Germany and UK, we focus on the anticyclonic AR regime for Germany and the cyclonic ZO regime for UK in the following. However, we note that results hold qualitatively for both countries (see Supplementary Dataset).  

\begin{figure}
    \centering
    \includegraphics[width=\linewidth]{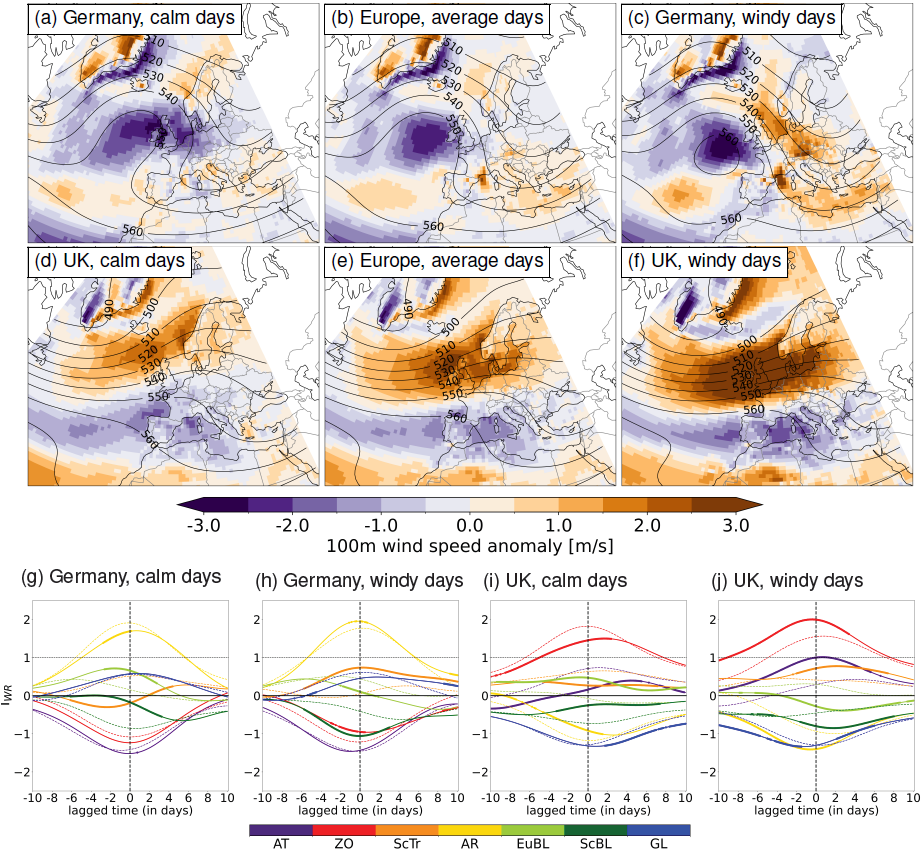}
    \caption{As in Figure \ref{fig:temp_example} but now for terciles of 100m wind speed anomalies in Germany during Atlantic Ridge events (a)-c),g),h)), and the UK during ZO events (d)-f),i),j).}
    \label{fig:wind_example}
\end{figure}

During calm AR days in Germany, the ridge over the North Atlantic extends further to the east and tilts in the southwest-northeast direction compared to average days (Figure \ref{fig:wind_example} a,b), indicative of anticyclonic wave breaking.  
Germany is located in a region of diffluent flow at 500\,hPa and the reduced geopotential height (and mean sea level pressure) gradients likely explain below average W100m anomalies.
The $I_{wr}$ of AR is slightly but significantly weaker than on average just prior to peak (Figure \ref{fig:wind_example} g). More importantly, calm AR days in Germany feature increased EuBL-$I_{wr}$ and decreased ScTr-$I_{wr}$ co-projections. 
Windy AR days in Germany feature a strong ridge over the North Atlantic with an average geopotential height anomaly of \SI{560}{gpdm} (Figure \ref{fig:wind_example} c), and a concomitant trough over Scandinavia. This leads to a strong pressure gradient over the North Sea and thus to very windy conditions over Germany. $I_{wr}$ for AR peaks around the windy days (Figure \ref{fig:wind_example} h), with a strong ScTr co-projection. Thus AR with concomitant EuBL and ScBL projections tends to calm conditions in Germany whereas a co-projection in ScTr is indicative of above average W100m anomalies during AR. 
Next we investigate ZO over the UK. During calm ZO days in UK, the ZO-$I_{wr}$ is below average (Figure \ref{fig:wind_example} i), and the stormtrack is slightly shifted north of the UK (cf. Figure \ref{fig:wind_example} d). Southern UK and part of the North Sea experiences 
 below-average wind speed. 

The area of windy anomalies is shifted to the north, smaller, and the anomalies are weaker compared to average ZO days (Figure \ref{fig:wind_example} d,e). 
At the same time co-projections into other cyclonic regimes are reduced, while $I_{wr}$ for EuBL is significantly enhanced (Figure \ref{fig:wind_example} i). 
The ZO days in the upper tercile of country-aggregated W100m anomalies for the UK are much windier than the average (Figure \ref{fig:wind_example} e,f), and the windy region extends to the Baltics and Germany. The area with below-average wind speeds is confined to the Mediterranean. 

Windy ZO days in UK occur during long, strong ZO events (Figure \ref{fig:wind_example} j) with co-projection into the other cyclonic regimes. Thus a co-projection in AT and ScTr is indicative of the expected windy conditions during ZO, while calm ZO day in UK project weaker in ZO and a co-projection in EuBL warrants caution with a likelihood of calmer conditions despite an active ZO regime.
These two examples show that wind speed anomalies are equally affected as temperature anomalies and can contain anomalies opposite to the mean anomaly. This is important to know for the wind energy sector since the cold and mainly calm regime AR can contain very windy days where the deficit in wind energy might be less than expected by the mean. Contrastingly, the windy-on-average ZO regime can be surprisingly calm. This work here presents which configuration of $I_{wr}$s favors a certain anomaly. Compared to the temperature anomalies, it shows another example of the importance of secondary regimes (AR), which can be found in many countries across variables and seasons (see Supplementary Dataset) and that the strength of the main regime can also be the most important change (ZO).

\section{Conclusions}
In this paper we have, for the first time, presented a detailed analysis of internal variability of surface weather within North Atlantic-European weather regimes, and provided a clear strategy for beginning to explain and account for that variability. 
With four different examples, we have shown that variations in surface weather within a particular regime can be, and often are, large, and so should not be ignored in applications. At the same time however, we show convincing evidence that much of this intra-regime variability can still be understood within a weather-regime framework, with the added aid of the continuous $I_{wr}$ indices. As such, we provide a path to more refined uses of weather regimes which allow a user to choose an acceptable trade-off between complexity and discriminatory power. Three main factors by which the $I_{wr}$ indices modify surface weather are identified: 1) Secondary regime indices can be anomalously high, indicating a 'hybrid' regime, 2) The primary $I_{wr}$ index may indicate that the surface anomalies preferentially occur early/late in a regime's life cycle, 3) the magnitude of the primary $I_{wr}$ index itself may be exceptionally strong or weak. Often a combination of at least two effects is visible. Rare cases do exist, where either no projection into the $I_{wr}$ is significant and/or there are no changes in the flow pattern, leaving room for further investigations of these special cases.
We believe these demonstrative results, supported by a comprehensive set of Supplementary Data \citep{Zenodo_2024}, will be of immediate use to the various stakeholders, e.g. in the energy or health sectors, and provide a refinement to the typical regime approach that will be of use to applied researchers. If we conceive of the multi-scale challenges of weather forecasting as requiring a `hierarchy of data-analysis' just as we acknowledge the value of a hierarchy of computational models, then the approach we describe provides a valuable middle-ground between a detailed (and data-intensive) consideration of the full atmospheric flow and a simple regime-mean framework.

\section{Open Research}
Information about data and code availability

\acknowledgments
The contribution of JD, and CMG was partially embedded in the subprojects A8 and T2 of the Transregional Collaborative Research Center SFB/TRR 165 ‘Waves to Weather’ (https://www.wavestoweather.de) funded by the Deutsche Forschungsgemeinschaft (DFG)". The contribution of MO was supported by Axpo Solutions AG. CMG acknowledges funding by the Helmholtz Association as part of the Young Investigator Group "Sub-seasonal Predictability: Understanding the Role of Diabatic Outflow" (SPREADOUT, grant VH-NG-1243). We thank Christopher Polster (Waves to Weather subproject A8 at University of Mainz) for providing the Flottplot package used in the supplementary dataset. We thank the authors of the python package "regionmask" for providing code via https://github.com/regionmask/regionmask used for country aggregation.   

\bibliography{main.bib}

\end{document}